\documentclass[aps,pra,reprint,twocolumn,floatfix,bibnotes,superscriptaddress]{revtex4-2} 
\usepackage{tabularx}
\usepackage{graphicx}
\usepackage{amsmath}
\usepackage{amssymb}
\usepackage{bm}
\usepackage{color}
\usepackage{dcolumn}
\usepackage{scalerel}
\usepackage{multirow}
\usepackage[caption=false]{subfig}
\usepackage{hyperref}
\usepackage{float}
\usepackage{dirtytalk}
\usepackage{makecell}
\usepackage{xcolor, soul}
\usepackage{orcidlink}

\newcommand{\NIST}{
National Institute of Standards and Technology, Boulder, Colorado 80305, USA
}

\newcommand{\CU}{
University of Colorado, Department of Physics, Boulder, Colorado 80309, USA
}

\newcommand{\infleqtion}{
present address: Infleqtion, 1315 West Century Drive, Suite 150, Louisville, Colorado 80027, USA
}

\newcommand{\EE}{
Electrical, Computer and Energy Engineering, University of Colorado, Boulder, Colorado 80309, USA
}

\definecolor{lime}{HTML}{A6CE39}
\DeclareRobustCommand{\orcidicon}{
	\begin{tikzpicture}
	\draw[lime, fill=lime] (0,0) 
	circle [radius=0.16] 
	node[white] {{\fontfamily{qag}\selectfont \tiny ID}};
	\draw[white, fill=white] (-0.0625,0.095) 
	circle [radius=0.007];
	\end{tikzpicture}
	\hspace{-0.3mm}
}

\foreach \x in {A, ..., Z}{\expandafter\xdef\csname orcid\x\endcsname{\noexpand\href{https://orcid.org/\csname orcidauthor\x\endcsname}
			{\noexpand\orcidicon}}
}

\makeatletter
\def\blfootnote{\gdef\@thefnmark{}\@footnotetext}
\makeatother

\begin{document}
\preprint{APS/123-QED}
\title{Two-photon Doppler-free spectroscopy of the $6s^2\;^1S_0 \rightarrow 6s23s\;^1S_0$ Rydberg transition for quantum sensing in ytterbium}

\author{Benjamin D. Hunt$\orcidC{}$}
\affiliation{\NIST}
\affiliation{\CU}
\author{Kyle Beloy$\orcidE{}$}
\affiliation{\NIST}
\author{Roger C. Brown$\orcidG{}$}
\affiliation{\NIST}
\author{Tobias Bothwell$\orcidI{}$} 
\altaffiliation{\infleqtion}
\affiliation{\NIST}
\author{Harikesh Ranganath$\orcidJ{}$} 
\affiliation{\NIST}
\affiliation{\CU}
\author{Tanner Grogan$\orcidK{}$} 
\affiliation{\NIST}
\affiliation{\CU}
\author{Jacob L. Siegel$\orcidB{}$}
\affiliation{\NIST}
\affiliation{\CU}
\author{Andrew D. Ludlow$\orcidH{}$}
\affiliation{\NIST}
\affiliation{\CU}
\affiliation{\EE}

\date{\today}

\begin{abstract}
Here we report atomic structure calculations and experimental observations of the two-photon $6s^2\;^1S_0 \rightarrow 6s23s\;^1S_0$ transition in Yb.  The unique level structure of Yb places the $6s6p\;^1P_1$ state's energy at $\approx\!49.7\%$ of the ionization energy, which  facilitates two-photon Doppler-free excitation of the $6s23s\;^1S_0$ Rydberg state from the $6s^2\;^1S_0$ ground state at a relatively low optical intensity. This $J= 0 \rightarrow J=0$ transition is immune to first-order Zeeman and electric quadrupole shifts, has reduced sensitivity to higher-order Zeeman shifts, and is further predicted to have relatively low sensitivity to density shifts. Using configuration interaction theory, we compute relevant properties of the $6s23s\;^1S_0$ Rydberg state for sensing, including the $C_6$ coefficient, dc polarizability, blackbody radiation (BBR) shift, and BBR depopulation rate. As a first experimental step, we demonstrate two-photon Doppler-free spectroscopy of the  $6s^2\;^1S_0 \rightarrow 6s23s\;^1S_0$ transition. From our experimental spectra, we estimate the natural linewidth (lifetime) of the $6s23s\;^1S_0$ state to be  90 kHz (1.8 $\mu$s).  We believe this transition, or other $6s^2\;^1S_0 \rightarrow 6sns\;^1S_0$ transitions in Yb with $n\approx23$, may find use in a variety of sensing applications, potentially including a two-photon clock,  electrometer, or BBR thermometer.
\end{abstract}

\maketitle

\section{\label{sec:intro} Introduction}


The comparatively large atomic polarizability and lifetime of Rydberg states have enabled precision electrometers, magnetometers, and thermometers ~\cite{sedlacek_microwave_2012,  fancher_rydberg_2021, bohaichuk_three-photon_2023, meyer_assessment_2020, teale_degenerate_2022, bowden_rydberg_2017,  dietsche_high-sensitivity_2019, norrgard_quantum_2021, mantia_2025,schlossberger_primary_2025, OvsDerGib11, fan2015atomRF}. An advantage of Rydberg-based quantum sensors is their ability to operate in warm atomic vapors, avoiding the need for laser cooling. However, in warm vapors, Doppler broadening and collisional effects can limit the achievable precision~\cite{holloway_electric_2017, meyer_2021, glick_warm_2025}. Alternative approaches to sensing based on cold atoms may increase precision at the cost of additional experimental complexity \cite{holloway_electric_2017, liao_2020, jamieson_2025, tu_2024}. While near-Doppler-free quantum sensing with Rydberg atoms has been demonstrated~\cite{bohaichuk_three-photon_2023,glick_warm_2025}, these implementations required either precise angular control of beam pointing~\cite{ryabtsev_doppler-_2011,glick_warm_2025} or frequency control of nondegenerate photons~\cite{bohaichuk_three-photon_2023} and have not completely eliminated first-order Doppler shifts. In contrast, degenerate two-photon Doppler-free Rydberg excitation might enable simplified sensors that maintain precision when operated in warm vapors~\cite{Lee78}.

The amplified blackbody radiation (BBR) sensitivity of Rydberg states~\cite{GalCoo79,FarWin81} makes them potentially attractive candidates for thermal sensing.  In Refs.~\cite{mantia_2025, schlossberger_primary_2025}, the population transfer rate of high-lying Rydberg states in Rb was used to determine the BBR environment with kelvin-level precision. One drawback of this approach is sensitivity to a narrow spectral component of the BBR, potentially introducing systematic error when used in non-Planckian environments.  In the proposal of Ref.~\cite{OvsDerGib11}, lattice-trapped Rydberg atoms were considered for sensing the thermal environment of atoms in an optical lattice clock~(OLC), allowing for in-situ characterization of the critical BBR Stark systematic shift.  Precise control of optical-lattice-induced light shifts of a Rydberg state can be a formidable challenge. On the other hand, a Doppler-free free-space thermometry approach would avoid the burden of lattice light shift control while offering high spectroscopic resolution. 

Here we explore the $6s^2\;^1S_0 \rightarrow 6s23s\;^1S_0$ two-photon transition both theoretically and experimentally. 
This transition possesses several advantageous properties for sensing: accessible degenerate two-photon Doppler-free excitation from the ground state, immunity to first-order Zeeman shifts and electric-quadrupole shifts, and reduced sensitivity to higher-order Zeeman shifts. As discussed below, sensors implementing the  $6s23s\;^1S_0$ state benefit from its $\approx 1.8$ $\mu$s natural lifetime, large dc polarizability, near-maximal sensitivity to BBR, a relatively suppressed $C_6$ coefficient, and lack of fine structure and hyperfine structure. In Sec.~\ref{sec: quantum sensing}, we briefly motivate the $6s^2\;^1S_0 \rightarrow 6s23s\;^1S_0$ transition for sensing applications. In Sec.~\ref{sec: experiment}, we demonstrate two-photon Doppler-free spectroscopy of the $6s^2\;^1S_0 \rightarrow 6s23s\;^1S_0$ transition, a practical step towards the larger goal of realizing quantum sensors with this transition. The Appendix contains tables of calculated energy levels, lifetimes, $C_6$ coefficients, dc polarizabilities, BBR shifts, BBR depopulation rates, and THz sensing frequencies. 

For discussion purposes, we assume a nuclear-spin-zero isotope of Yb (e.g., $^{174}$Yb, which is the most abundant isotope). However, for the experimental work described in Sec.~\ref{sec: experiment}, the nuclear-spin-1/2 isotope $^{171}$Yb is used instead. This is done as an act of convenience, since our existing optical lattice clock apparatus typically operates with this isotope. Consequences of this choice are addressed in Sec.~\ref{sec: experiment}. Unless otherwise stated, we assume room-temperature BBR. Physical constants and symbols appearing in expressions below include the principal quantum number and effective principal quantum number of the Rydberg electron ($n$ and $n^*$, respectively), the speed of light ($c$), Planck's constant ($h$), Planck's reduced constant ($\hbar$), Boltzmann's constant ($k_B$), the elementary charge ($e$), the permittivity of free space ($\epsilon_0$), the mass of the electron ($m_e$), the Bohr radius ($a_B$), and the fine structure constant ($\alpha$).

\section{Quantum sensing with the $6s^2\;^1S_0 \rightarrow 6s23s\;^1S_0$ Transition}
\label{sec: quantum sensing}
We begin with some general atomic structure and experimental considerations, which motivate our selection of a particular Rydberg state.  For quantum sensing based on laser spectroscopy, first-order Doppler effects are an immediate concern. This includes both broadening and shifts to the spectral line, potentially compromising statistical and systematic uncertainty. These effects, as well as atomic recoil shifts, can be suppressed or eliminated with two-photon spectroscopy using counterpropagating probe beams~\cite{Biraben_2photonreview_2019,Goeppert_1931,Goeppert_1931_translated}. For good suppression, the frequencies of the counterpropagating beams should be nearly equal. Meanwhile, to realize practical transition rates, the beams should have small detuning with respect to electric-dipole-allowed transitions connecting the initial and final states via an intermediate state. These criteria generally limit the applicability of two-photon Doppler-free spectroscopy.

The $6s^2\;^1S_0 \rightarrow 6s6p\;^1P_1$ transition is commonly used as a cooling and read-out transition in ytterbium experiments. Energetically, the $6s6p\;^1P_1$ state lies anomalously near halfway ($49.7\%$) between the $6s^2\;^1S_0$ ground state and the first ionization threshold, corresponding to a Yb$^+$ ion in the $6\,^2S_{1/2}$ ground state~\cite{NIST_ASD}. Noting that the Rydberg states lie just below the ionization threshold in the energy spectrum, this opens up the possibility for two-photon spectroscopy from the ground state to a Rydberg state, with the $6s6p\;^1P_1$ state serving as an intermediate state. To look more closely at a possible laser excitation scheme, we consider only states that satisfy total-electron-spin and single-electron-jump selection rules from the $6s6p\;^1P_1$ intermediate state. Specifically, this limits us to the $6sns\;^1S_0$ and $6snd\;^1D_2$ Rydberg series. Assuming degenerate (i.e., equal-frequency) excitation photons, the energy of the Rydberg state determines the detuning of the virtual state with respect to the $6s6p\;^1P_1$ intermediate state.

\begin{figure}[t]
\includegraphics[width=246pt]{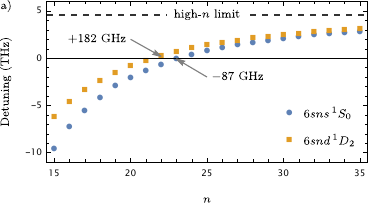}
\vspace{1mm}\\
\includegraphics[width=246pt]{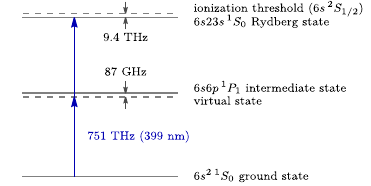}
\caption{Two-photon excitation of Rydberg states in Yb, starting from the $6s^2\,{^1S_0}$ ground state. Degenerate excitation photons are assumed. (a) Detuning from the $6s6p\,{^1P_1}$ intermediate state for excitation to $6sns\,{^1S_0}$ and $6snd\,{^1D_2}$ Rydberg states, for $15\leq n\leq 35$. Negative (positive) values correspond to red (blue) detuning with respect to the $6s^2\,{^1S_0}\rightarrow6s6p\,{^1P_1}$ transition. The dashed line denotes the high-$n$ limit common to both series. For each series, the state with the smallest detuning is highlighted with arrows and the detuning is shown. (b) Depiction of two-photon excitation to the $6s23s\,{^1S_0}$ Rydberg state. For visual clarity, the small frequency intervals are exaggerated relative to the photon frequency ($4\times$ in the case of the 9.4~THz interval, $400\times$ in the case of the 87~GHz interval).}
\label{Fig:levelsplot}
\end{figure}

In Figure ~\ref{Fig:levelsplot}(a), we plot this detuning for both series, borrowing Rydberg state energies from Ref.~\cite{JohBluSap88}. The plot spans the range $15 \leq n \leq  35$, where the detunings exhibit a transition from negative to positive values. Here negative (positive) values correspond to a photon frequency red (blue) detuned with respect to the $6s^2\;^1S_0 \rightarrow  6s6p\;^1P_1$ transition. For the $6sns\;^1S_0$ series, the smallest detuning is found to be $-87$ GHz, attributed to the $6s23s\;^1S_0$ state. For the $6snd\;^1D_2$ series, the smallest detuning is found to be $+182$ GHz, attributed to the $6s22d\;^1D_2$ state.

Aside from offering a smaller detuning for degenerate photons, the $6sns\;^1S_0$ series offers additional advantages over the $6snd\;^1D_2$ series. The $6sns\;^1S_0$ states are nondegenerate. They are immune to first-order Zeeman and electric quadrupole shifts. They are not part of a fine structure manifold, with higher-order Zeeman shifts consequently being suppressed. The lack of structure also improves the signal-to-noise ratio in situations where state preparation is not feasible. For thermal sensing applications, due to their spherical symmetry, they sample the BBR environment without directional bias. Background signals attributed to the absorption of two photons from a single beam direction can also be readily suppressed with an appropriate choice of laser polarization. Namely, using circular polarization with like-helicity for the counterpropagating beams, only transitions associated with absorption of a photon from each beam are allowed. We show the two-photon $6s^2\;^1S_0 \rightarrow 6s23s\;^1S_0$ transition in Fig.~\ref{Fig:levelsplot}(b).

Quantum sensors must balance the need to provide a strong response to wanted signals while remaining insensitive to unwanted noise sources or biases.  Higher-lying Rydberg states are often seen as favorable, given their increased dc polarizabilities ($\alpha_{\text{dc}}$) that scale as $(n^*)^7$ within a given series, as well as the potential for improved spectroscopic precision from longer natural lifetimes ($\tau_\mathrm{nat}$) that scale as $(n^*)^3$.  However, for some applications these higher-lying states suffer from increased sensitivity to density-dependent van der Waals interactions ($C_6$) that scale much more steeply with $n^*$, as $(n^*)^{11}$~\cite{Kitching25}. For the applications we discuss in Sec.~\ref{sec: quantum sensing} B, we choose the $6s23s\;^1S_0$ state in part because it strikes a balance between these considerations.

\begin{figure}[h]
\includegraphics[width=246pt]{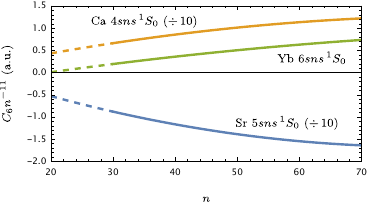}
\caption{$C_6$ coefficients for the $6sns\,{^1S_0}$ series of Yb from Vaillant~{\it et al.}~\cite{VaiJonPot12}. $C_6$ coefficients for the analogous ${^1S_0}$ series in Ca and Sr are also included to illustrate the relative suppression for the Yb case, with the Ca and Sr results being divided by a factor of 10 in order to be displayed on a similar scale. The curves are polynomial functions of $n$ given in Ref.~\cite{VaiJonPot12}, which are fits to $C_6$ coefficients evaluated over the range $30\leq n\leq70$. The dashed portion of the curves denote extrapolation outside of the evaluated range. The extrapolated Yb curve crosses zero at $n=19.6$. The $C_6$ coefficients are in atomic units (a.u.), with the dominant $\sim\!n^{11}$ scaling factored off.}
\label{Fig:vaillant}
\end{figure}

Considering atomic interactions more specifically, Vaillant {\it et al.}\ computed $C_6$ coefficients for various Rydberg series of the alkaline earth(-like) systems Ca, Sr, and Yb ~\cite{VaiJonPot12}. A semi-empirical method was employed, using experimental energies as input. The method is based on a single-active-electron model, with the Coulomb approximation being used to determine electric dipole matrix elements. Comparing results for the different series, the $C_6$ coefficients for the $6sns\;^1S_0$ series of Yb are seen to be relatively suppressed. The $6sns\;^3S_1$ series of Yb was not considered in Ref.~\cite{VaiJonPot12}, but the $C_6$ coefficient for the $6s50s\;^3S_1$ state was recently measured in an optical tweezer experiment~\cite{MaBurLiu22}. The result is $\approx40$ times greater than the theoretical value for the $6s50s\;^1S_0$ state given in Ref.~\cite{VaiJonPot12}, further exemplifying the relative suppression for the $6sns\;^1S_0$ series. Unfortunately, the states presently of interest ($n \approx 23$) lie outside the range of states evaluated in Ref.~\cite{VaiJonPot12}, which was limited to $30 \leq  n \leq  70$. However, Ref.~\cite{VaiJonPot12} does provide a simple polynomial function of $n$, intended to encapsulate the $C_6$ coefficients over the evaluated range. Extrapolating to lower values of $n$, the value of this function decreases until crossing zero at $n = 19.6$. This hints at possible additional suppression for the $6sns\;^1S_0$ states in the region of interest. We illustrate this suppression in Fig.~\ref{Fig:vaillant}, while stressing that any conclusions drawn from the extrapolation are merely speculative.

\subsection{Configuration interaction calculations of the $ 6s23s\;^1S_0$ state}
\label{sec: ci_theory}

Given that the $6s23s\;^1S_0$ state is not exceptionally high in the energy spectrum,  computational methods typically applied to low-lying atomic states are potentially applicable here. In Ref.~\cite{GolNorKol15}, for instance, properties of the $18{\;}^2S_{1/2}$ Rydberg state of Rb were evaluated using an all-order technique developed primarily for low-lying atomic states of monovalent systems, with special considerations being made in the selection of computational parameters. In the present work, we employ a configuration interaction (CI) method, treating Yb as a divalent system. The method, originally developed to evaluate properties of low-lying states of divalent systems~\cite{CheCheJoh93,SavJoh02}, is detailed in the Appendix.  We use it to calculate a number of atomic properties relevant to quantum sensing with the $6s23s\;^1S_0$ state, including the $C_6$ coefficient, dc polarizability ($\alpha_{\text{dc}}$), second-order Zeeman shift, BBR shift ($\delta E_{\text{BBR}})$, BBR sensitivity ($S_{\text{BBR}}$), and BBR depopulation rate ($\gamma_{\text{BBR}}$) of this state. These results are compiled in Table~\ref{Tab:transition_props}. In this table, we also include the state energy, the estimated two-photon Rabi frequency ($\Omega$), and the ac probe Stark shift $(\Delta\nu_{\text{ac}}$). These properties and parameters are considered in various quantum sensing applications below. The CI calculations support a suppressed $C_6$ coefficient, consistent with extrapolated results from Ref.~\cite{VaiJonPot12} (see the Appendix for more details). The second-order Zeeman shift for this transition is dominated by the Rydberg state, which itself is estimated to be dominated by the diamagnetic contribution. For this diamagnetic contribution, we find agreement at the $\approx 0.3\%$ level between the CI result and the result obtained from a simple hydrogenic formula using the effective principal quantum number of the Rydberg electron~\cite{beloy_2025}.

We further use the CI atomic structure calculations to investigate possible detection channels for a quantum sensor using the $6s23s\;^1S_0$ state. The largest branching factors from the $6s23s\;^1S_0$ state are to the $6s6p\;^1P_1$ state (81\%), the $6s7p\;^1P_1$ state (4.7\%), the $6s8p\;^1P_1$ state (3.0\%), the $6s9p\;^1P_1$ state (1.5\%), and the $6s6p\;^3P_1$ state (1.1\%). However, these results do not account for configuration mixing with states having a $4f$-excited configuration, as discussed in the Appendix. Such mixing opens up decay to these states as well, most notably the $4f^{13}5d6s^2$ state lying immediately above the $6s6p\;^1P_1$ state in the $J^\pi=1^-$ energy spectrum (see Table~\ref{Tab:Energies1o} of the Appendix; $\pi$ denotes parity). A rudimentary estimate based on state energies and lifetimes indicates that the resulting branching to this state could be as large as $\sim\!10\%$, with a corresponding decrease in branching to the $6s6p\;^1P_1$ state. 470~nm and 347~nm photons are emitted as the atom cascades from the $6s23s\;^1S_0$ state to the ground state via this $4f$-excited state. We describe three distinct experimentally realized detection channels from $6s23s\;^1S_0$ decay in Sec.~\ref{sec: experiment}.

\begin{table}[t]
\caption{Properties of the $6s23s\;^1S_0$ Rydberg state and the  $6s^2\,^1S_0 \rightarrow 6s23s\,^1S_0$ two-photon transition in Yb. Here, $I_1$ and $I_2$ are the intensities of the individual counterpropagating excitation beams and degenerate photons are assumed. $\gamma_\mathrm{nat}/2\pi$ is the natural linewidth.}
\label{Tab:transition_props}
\begin{ruledtabular}
\begin{tabular}{l c}
\textbf{Quantity} & \textbf{Value} \\
\hline
Energy (cm$^{-1}$)\footnotemark[1] & 50130.6 \\
Second-order Zeeman (Hz/G$^2$)\footnotemark[2] & 30.7 \\
$\dfrac{\Omega/(2\pi)}{\sqrt{I_1 I_2}}$ (Hz/(mW/mm$^2$))\footnotemark[2] & 330 \\
$\dfrac{\Delta\nu_{\text{ac}}}{I_1 + I_2}$ (Hz/(mW/mm$^2$))\footnotemark[2] & 2000 \\
$C_6$ ($\times 10^{13}$ a.u.)\footnotemark[2] & 7.6 \\
$\alpha_{\text{dc}}$ (Hz/(V/m)$^2$)\footnotemark[2] & 77 \\
$\gamma_\mathrm{nat}/2\pi$ (kHz)\footnotemark[3] & 90 \\
$\delta E_{\text{BBR}}/h$ (kHz)\footnotemark[2] & 3.0 \\
$S_{\text{BBR}}/h$ (Hz/K)\footnotemark[2] & 18 \\
$\gamma_{\text{BBR}}$ (ms$^{-1}$)\footnotemark[2] & 36 \\
\end{tabular}
\end{ruledtabular}

\footnotetext[1]{Ref.~\cite{LehZulMai18}}
\footnotetext[2]{modified CI; see Appendix}
\footnotetext[3]{Sec.~\ref{sec: experiment}}
\end{table}

\subsection{Sensing Applications}
\label{Sec:applications}

{\it Two-photon clock.} The $6s^2\;^1S_0 \rightarrow 6s23s\;^1S_0$ transition possesses a number of similar features to the well-established two-photon (778 nm) clock operated on the $5\;S_{1/2} \rightarrow 5\;D_{5/2}$ transition in Rb with 330 kHz natural linewidth~\cite{NEZ1993Rb}. The $5S_{1/2} \rightarrow 5D_{5/2}$ Rb transition has a $\approx\!1$ THz intermediate-state detuning for Doppler-free excitation with reasonable intensities~\cite{beard_two-photon_2024, lemke_measurement_2022, martin_compact_2018, ahern_tailoring_2025, poulin_frequency_2002}, no first-order Zeeman shift, and possesses a 7\% decay channel at 420~nm for background-free fluorescence detection. The Yb transition shares each of these features, with a comparable fluorescence branch near $470$~nm, but with reduced second-order Zeeman shifts, intermediate-state detuning $\approx\!87$~GHz, as well as a narrower natural linewidth, and higher clock frequency.  To speculate on the performance potential, we note a recent demonstration of an ytterbium vapor clock using the $2\times$ broader and $3\times$ lower frequency $6s^2\;^1S_0 \rightarrow 6s6p\;^3P_1$ transition, which achieved a one-second fractional instability of $\approx\!4\times 10^{-14}$~\cite{hilton_demonstration_2025}.

{\it Rydberg electrometry.} The dc Stark shift for a non-degenerate state can be expressed as $\Delta E_{\text{dc}} = -\frac{1}{2} {\alpha_{\text{dc}}}E^2$
where $E$ is an applied electric field.  When placed in a parallel plate geometry, this electric field sensitivity may be used as a voltage reference~\cite{teale_degenerate_2022,SheLemHin12, jau_2020}. For specificity, we compare the atomic properties of a hypothetical voltage reference using the $6s^2\;^1S_0 \rightarrow 6s23s\;^1S_0$ transition to an analogous Doppler-free two-photon voltage reference in Ref.~\cite{teale_degenerate_2022} operated using the $6S_{1/2} \rightarrow 15S_{1/2}$ transition in Cs. We find that the $6s23s\;^1S_0$ state has a $\approx\!150 \times$ larger $\alpha_{\text{dc}}$ (see Table~\ref{Tab:transition_props}, with only a $3\times$ decrease in the accessible voltage range for a given electrode spacing) \footnote{The $6sns\;^1S_0$ states with $n=14$ to $n=26$ described in the Appendix may also be suitable alternatives to the $6s23s\;^1S_0$ state depending on the desired sensitivity and range}.
The Yb natural linewidth observed in this work is $\approx 50\times$ smaller than in Ref.~\cite{teale_degenerate_2022}.
Additionally, the $J=0$ nature of the Yb $6s23s\;^1S_0$ state eliminates $E$-field-dependent broadening and other systematic effects typically present in $J>0$ Rydberg series. 
With the strength of the Cs and Yb Rydberg dipole matrix elements being a similar order of magnitude, the reduction of the intermediate-state detuning from $\approx\!100$~THz in Cs to $\approx 100$~GHz in Yb should result in a $>\!100\times$ reduction in required laser power for a given two-photon Rabi frequency.     

Rydberg-atom rf electrometers have realized highly sensitive, SI-traceable measurements and imaging of radio-frequency electric fields in atomic vapors~\cite{sedlacek_microwave_2012, fancher_rydberg_2021, bohaichuk_three-photon_2023, meyer_assessment_2020, holloway_electric_2017, meyer_2021}. The $6s^2\;^1S_0 \rightarrow 6s23s\;^1S_0$ transition might be useful for sensing terahertz frequencies~\cite{wade_real-time_2017}. In the Appendix we include a table of the reduced electric-dipole matrix elements and corresponding transition energies between the $6s23s\;{}^1S_0$ state and selected members of the $6snp~J=1$ series with $n\approx 23$.

{\it Blackbody radiation thermometry.} In the high-$n$ limit, a simple universal expression for the Stark shift due to BBR is predicted to hold~\cite{FarWin81},
$\delta E_\mathrm{BBR}\approx
\frac{\pi\alpha\left(k_BT\right)^2}{3m_ec^2}$, 
where $T$ is the BBR temperature. This expression evaluates to $\delta E_\mathrm{BBR}/h\approx2.4~$kHz at $T=300$~K. $\delta E_{\text{BBR}}$ has been calculated for various Rydberg series in multiple atomic species~\cite{FarWin81,OvsDerGib11}. For all Rydberg series considered, a tendency towards the high-$n$ limit is observed with increasing $n$, though the progression towards this limit generally does not exhibit simple behavior. Broadly speaking, however, the high-$n$ limit is seen to be approximately satisfied already by $n\approx20$. Therefore, with regard to BBR sensitivity, Rydberg states with higher values of $n$ generally do not offer meaningful advantage in sensitivity, but may rather incur steep systematic effect penalties.

The BBR temperature can potentially be extracted with an uncertainty $\sigma_T\approx(60~\text{mK/Hz})\sigma_\nu$, where $\sigma_\nu$ denotes measurement uncertainty in the frequency of the transition from a low-lying state to the Rydberg state used for BBR sensing. In practice, both statistical and systematic contributions to $\sigma_\nu$ must be taken into account. For a BBR thermometer operating on the $6s^2\:^1S_0 \rightarrow 6s23s\;^1S_0$ transition, the primary anticipated systematic stems from the probe ac Stark shift $\Delta\nu_{\text{ac}}$. Based on the BBR sensitivity and natural linewidth for this transition, a 1~K temperature resolution requires resolving the line center relative to the linewidth (``splitting the line'') by $\approx10^4$. Unlike Ref.~\cite{OvsDerGib11}, here no magic wavelength trapping is required for precision measurement of the transition frequency. While spectroscopic line splitting has been realized at finer levels in microwave fountain clocks (with Fourier-limited linewidths), one part in $10^4$ remains ambitious. To extract accurate temperature measurements, an atomic ``calibration'' of the atomic frequency in a well-characterized BBR environment would be required. The resulting calibration curve could then be applied to systems with more relaxed environmental control, possibly including deviations from a Planckian spectral distribution. Implications of these deviations would generally need to be taken into account, as highlighted in the Appendix.

\section{Proof of Principle Two-Photon Doppler-Free Spectroscopy of the $6s^2\;^1S_0 \rightarrow 6s23s\;^1S_0$ Transition}
\label{sec: experiment}

Here we describe three proof-of-principle spectroscopy demonstrations of the $6s^2\;^1S_0\rightarrow6s23s\;^1S_0$ transition in Yb, each using distinct detection methods as described below. As discussed in the Introduction, we use the nuclear-spin-1/2 isotope $^{171}$Yb rather than a nuclear-spin-zero isotope as an act of convenience. We briefly discuss consequences of this choice. Below $F$ and $m_F$ identify with the conventional total angular momentum quantum numbers.

For the nuclear-spin-zero isotopes, the $6sns\;^1S_0$ series is well-defined for arbitrary $n$. For isotopes with nonzero nuclear spin, this is not the case. For large $n$, the hyperfine interaction in these isotopes causes the $6sns\;^1S_0 \;F=1/2$ and $6sns\;^3S_1\;F=1/2$ series to be strongly mixed, with $J$ no longer being a good quantum number~\cite{peper2025}. For $n=23$, the mixing is still relatively small. We estimate a residual linear Zeeman shift of $\approx\!170$~Hz/G due to this hyperfine mixing, which is not observable in the present experiments.

For the nuclear-spin-zero isotopes, background signals due to absorption of two photons from a single beam are eliminated when using circularly polarized probing light. This is because the absorption of two circularly polarized photons from a single beam requires a change of the $m_F$ quantum number by two units. For nuclear-spin-zero isotopes, $J=0$ for both states implies $F=0$ for both states, and it follows that $m_F$ cannot change by two units. Conveniently, this background elimination also works for the nuclear-spin-1/2 isotope $^{171}$Yb. In this case, $F=1/2$ for both states, and it similarly follows that $m_F$ cannot change by two units.

\begin{figure}[t]
\includegraphics[width=246pt]{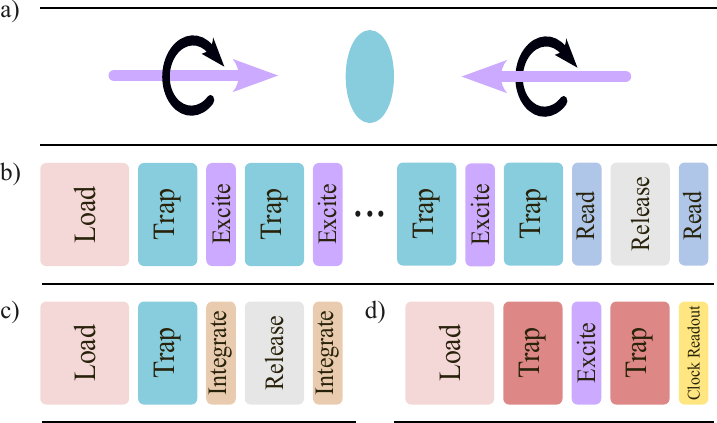}
\caption{Experimental protocol for two-photon Doppler-free spectroscopy of the $6s^2\;^1S_0\rightarrow6s23s\;^1S_0$ transition. (a) Illustration of the two-photon excitation geometry. Counterpropagating $\approx\!399$ nm photons with circular polarization of like-helicity (purple arrows) are used to excite free-space atoms (teal oval) from the $6s^2\;^1S_0$ ground state to the $6s23s\;^1S_0$ state. (b) Simplified timing diagram of trap-loss spectroscopy (time axis not to scale for clarity). Atoms are loaded into a 3D MOT before 5~ms Rydberg excitation pulses are time-interleaved with 20~ms MOT recapture periods. MOT population is measured with MOT fluorescence collected by an Electron Multiplying Charge-Coupled Device (EMCCD). Background fluorescence is corrected for with an additional fluorescence measurement after the MOT is extinguished. (c) Timing diagram used for fluorescence spectroscopy of the $6s^2 \; ^1S_0 \rightarrow 6s23s \; ^1S_0$ transition. A green (blue) MOT is used to capture atoms before the MOT is extinguished and the 399 nm (470 nm) fluorescent photons are integrated on an  EMCCD for the entirety of the 5~ms spectroscopy pulse before the background is corrected. (d) Timing diagram used for lattice-trapped $6s6p \;^3P_0$ population spectroscopy of the $6s^2 \; ^1S_0 \rightarrow 6s23s \; ^1S_0$ transition. Atoms are loaded from a green MOT into a lattice of trap depth 100$E_r$, where $E_r = \frac{\hbar^2 k^2}{2m}$ represents the recoil energy of $^{171}$Yb of mass $m$ from a 759 nm photon with wavenumber $k$, before the lattice is extinguished for $1$ ms and the now-unconfined atoms are excited on the $6s^2 \; ^1S_0 \rightarrow 6s23s \; ^1S_0$ transition. Following excitation, the lattice is restored and the atomic population in the $6s^2\;^1S_0$ and $6s6p\;^3P_0$ states is measured as described in Refs.~\cite{McGZhaFas18, chen_2024}.}
\label{Fig:sequence}
\end{figure}

In our OLC, an atomic beam is produced from a metallic Yb sample heated to $\approx\!450~^{\circ}$C. $^{171}$Yb atoms are slowed on the ``blue" $6s^2\;^1S_0\rightarrow 6s6p\;^1P_1$ transition at $399$ nm before they are loaded into a blue magneto-optical trap (MOT). For the lattice-trapped experiments described later, atoms are further cooled through four $556$~nm ``green'' MOT stages operated on the $6s^2\;^1S_0\rightarrow 6s6p\;^3P_1$ transition, before being loaded into an optical lattice operated near the $759$~nm magic wavelength for the $6s^2 \;^1S_0$ and $6s6p \; ^3P_0$ clock states~\cite{bothwell_2025}. The setup we use for spectroscopy of the $6s^2\;^1S_0\rightarrow 6s23s\;^1S_0$ transition is shown in Fig.~\ref{Fig:sequence}(a).

Our spectroscopy laser is a 399~nm commercial external cavity diode laser that we frequency stabilize to an isolated optical cavity with finesse $\mathcal{F} \approx 1500$ by Pound-Drever-Hall locking. A double-passed acousto-optic modulator in the cavity path is used to set the spectroscopy laser frequency. We determine the energy of the $6s23s\,^1S_0$ state in $^{171}$Yb to be 50130.6710(7)~cm$^{-1}$, limited by wavemeter accuracy, and observe sub-Doppler features with three detection techniques: trap-loss spectroscopy, detection of lattice-trapped atoms accumulated in the $6s6p\;^3P_0$ metastable state, and fluorescence spectroscopy. 

\begin{figure*}[t]
    \centering
    \includegraphics[width=\textwidth]{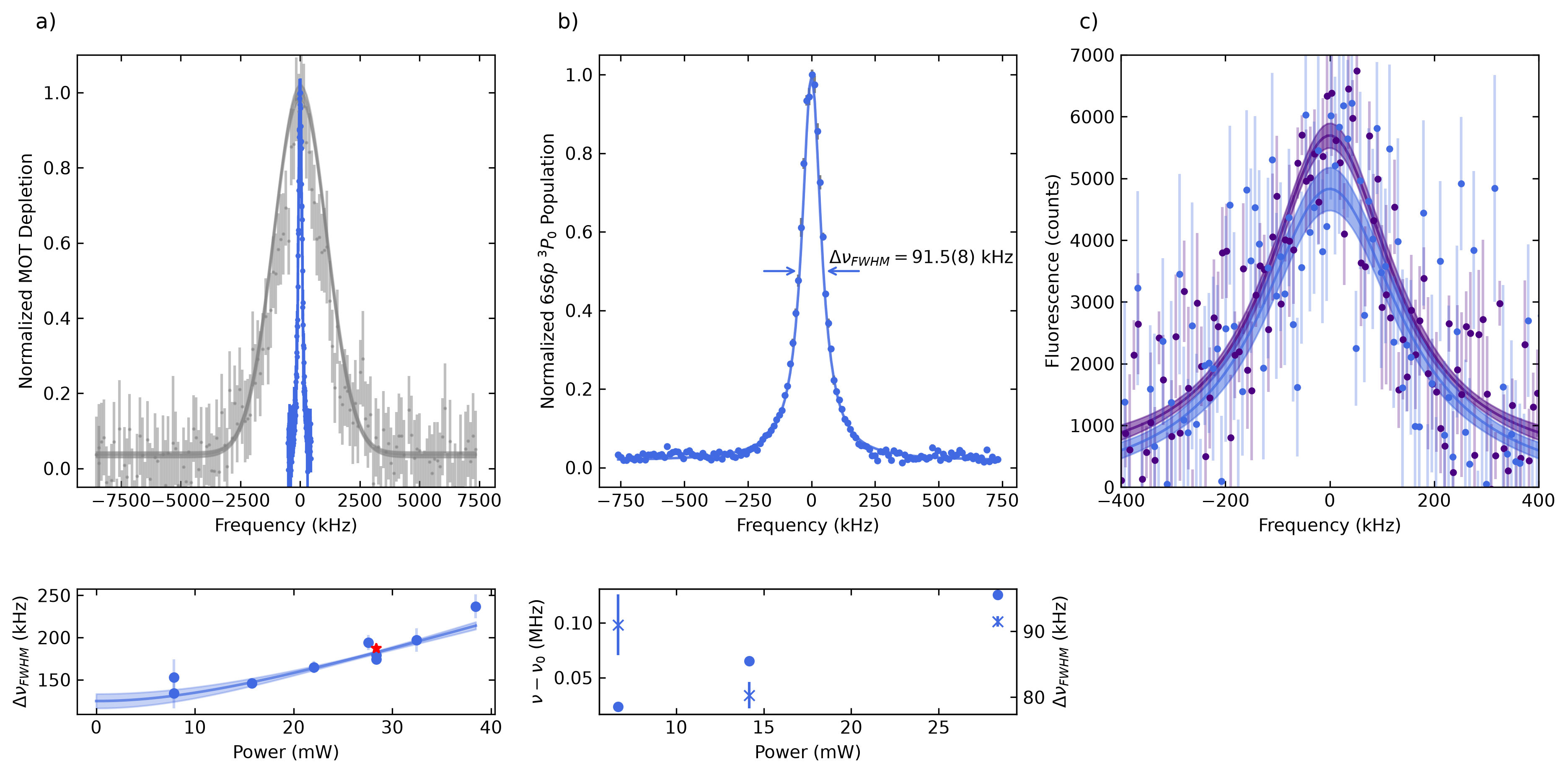}
    \caption{Experimental spectra for the $6s^2\;^1S_0\rightarrow6s23s\;^1S_0$ two-photon transition. (a, upper panel) Doppler-broadened (gray data points) and Doppler-free (blue data points) spectra obtained using trap-loss spectroscopy as described in the main text. The frequency axis is relative to the fitted center frequency of the transition. The solid gray curve is a Gaussian fit to the Doppler-broadened data, while the solid blue curve is a Lorentzian fit to the Doppler-free data. Shaded regions give the respective $1\sigma$ statistical uncertainty. (a, lower panel) Linewidth as a function of total laser power ($P \propto I_1 + I_2$). The solid curve is the fitted approximate linewidth of a Voigt profile, Eq.~(\ref{eq:voigt}), with shaded $1\sigma$ uncertainty. The red star is the Doppler-free measurement in the upper panel. (b, upper panel) Doppler-free profile from lattice-trapped $6s6p\;^3P_0$ population spectroscopy (data points). The blue solid curve is a Lorentzian fit with shaded $1\sigma$ uncertainty. (b, lower panel) ac Stark shift (circle markers, left axis) and $\Delta \nu_{\mathrm{FWHM}}$ ($\times$ markers, right axis). Here $\nu_0$ corresponds to the estimated unperturbed transition frequency. (c) Background-corrected integrated fluorescence spectra, detecting 399~nm photons (violet data points) and 470~nm photons (blue data points) as described in the main text. Solid curves of the respective color are Lorentzian fits with shaded $1\sigma$ uncertainty.}
    \label{fig:combined}
\end{figure*}

Trap-loss spectroscopy, where Rydberg excitation is detected by a decrease in MOT population, is a well-established spectroscopy technique for Rydberg states and has been performed with multiple species~\cite{couturier_2019, hostetter_2015, Moi_2013, Hou2025, duverger_2024}. We perform our trap-loss spectroscopy of the $6s^2\;^1S_0 \rightarrow 6s23s\;^1S_0$ transition in a blue MOT. To reduce trap-induced broadening, Rydberg excitation is time-interleaved with the trapping light as illustrated in Fig.~\ref{Fig:sequence}(b). For each detuning of the spectroscopy laser, 15 to 50 cycles of interleaved depletion are repeated before the MOT population is measured. During the interleaved sequence, the MOT quadrupole field is held constant at $\approx10 $ G$/$cm. To account for other loss mechanisms, the atomic population in the Rydberg-depleted MOT is normalized to the population of an otherwise identical MOT with no excitation.


 The upper panel of Fig.~\ref{fig:combined}(a) shows a sub-Doppler spectrum measured with counterpropagating circularly polarized photons of like-helicity. For comparison, we also include a Doppler-broadened spectrum obtained from a single linearly polarized beam. The full-width-at-half-maximum (FWHM) linewidth of the Doppler-broadened feature is 2.7(7)~MHz, corresponding to a mean atomic temperature $T \approx 1$~mK. The FWHM linewidth of the Doppler-free feature is 187(3)~kHz, corresponding to a $14\times$ reduction compared to the Doppler-broadened profile. For each measurement presented in Fig.~\ref{fig:combined}(a), the total depletion is limited to $<\!50\%$ of the initial atom number to avoid saturation effects. We observe no measurable Zeeman broadening from MOT field gradients from 2~G/cm to 10~G/cm. We observe broadening that is dependent on spectroscopy laser power. The lower panel of Fig.~\ref{fig:combined}(a) shows the measured linewidth versus spectroscopy laser power. We assume the following formula for the linewidth as a function of laser power,
\begin{equation}
    \Delta \nu_{\mathrm{FWHM}}(P) = b_0 \Delta \nu_l + \sqrt{\; b_1 \Delta\nu_l^2 + (\delta\nu_g P)^2}, 
\label{eq:voigt}
\end{equation}
where $b_0=0.5343$ and $b_1=0.2169$ are fixed parameters and $P$ is the laser power. This formula closely approximates the FWHM of a Voigt profile with a power-independent Lorentzian component  of FWHM $\Delta\nu_l$ and a linear-in-power Gaussian component of FWHM $\delta\nu_gP$. Fitting $\Delta\nu_l$ and $\delta\nu_g$ to the data of Fig.~\ref{fig:combined}(a) yields $\Delta\nu_l= 125(8)$~kHz and $\delta\nu_g = 3.5(3)$~kHz/mW, with $\chi^2_{\mathrm{red}}=1.8$.
We attribute the Gaussian component to an inhomogeneous ac Stark shift stemming from the spatial profile of the spectroscopy laser ($\approx 1$\;mm) sampled by atoms released from the blue MOT ($\approx 1$\;mm).



We observe atomic population in the metastable $6s6p \; ^3P_0$ clock state following excitation of atoms to the $6s23s\;^1S_0$ state. This enables spectroscopy of the $6s^2\;^1S_0 \rightarrow 6s23s\;^1S_0$ transition using $6s6p \; ^3P_0$ population as a detection channel. The experimental sequence for spectroscopy using this detection method is shown in Fig.~\ref{Fig:sequence}(d). Using this method, the upper panel of Fig.~\ref{fig:combined}(b) shows a measured spectrum with linewidth of 91.5(8)~kHz. With additional data presented in the lower panel of Fig.~\ref{fig:combined}(b), we estimate the average measured linewidth (lifetime) of the $6s23s\;^1S_0$ state to be 90~kHz (1.8~$\mu$s). To our knowledge, this is the narrowest optical spectral feature resolved for a transition to a Rydberg state, realized with a simple degenerate two-photon scheme and outperforming even three-photon strategies to date~\cite{bohaichuk_three-photon_2023}. We attribute the disagreement between this linewidth estimation and the $\Delta\nu_l$  extrapolated from trap-loss spectroscopy to broadening effects from residual blue MOT trapping light, as well as possible Zeeman or interatomic effects, or other unknown broadening mechanisms present in the trap-loss experiments. With the lattice-trapped $6s6p\; ^3P_0$ population detection method, we measure the two-photon ac Stark shift of the $6s^2\;^1S_0\rightarrow6s23s\;^1S_0$ transition to be $\approx\!2$~kHz/mW as shown in the lower panel of Fig.~\ref{fig:combined}(b). Combining this with the theoretical ac Stark shift in Table~\ref{Tab:transition_props}, we estimate the spectroscopy beam to have an $\approx 1.0$ mm $1/e^2$ intensity diameter.

We further perform fluorescence spectroscopy of the $6s^2 \; ^1S_0 \rightarrow 6s23s \; ^1S_0$ transition with the experimental sequence shown in Fig.~\ref{Fig:sequence}(c), collecting either 399~nm or 470~nm fluorescence. Two 399~nm photons are emitted in the cascaded decay $6s23s\;^1S_0 \rightarrow 6s6p\;^1P_1 \rightarrow 6s^2\;^1S_0$, while a single 470~nm photon is emitted as the $6s23s\;^1S_0$ state decays to the $4f$-excited state discussed in Sec.~\ref{sec: ci_theory}.
We filter decay photons from background radiation with a bandpass filter centered at either 400~nm or 470~nm. In the respective cases, we measure linewidths of 255(23)~kHz and 352(50)~kHz, with the measured spectra shown in Fig.~\ref{fig:combined}(c). These measurements were taken with the probe power at $\approx30$~mW and no extrapolation to zero power. We attribute the relative broadening of these spectral lines to the same inhomogeneous light shift across the atomic ensemble as described for the trap-loss spectroscopy.  
 

\section{Conclusion}
Here we have proposed using the $6s^2\;^1S_0\rightarrow6s23s\;^1S_0$ two-photon transition in Yb as a two-photon clock, voltage reference, or free-space BBR thermometer. This two-photon transition is immune to first-order Zeeman shifts and electric quadrupole shifts, and has reduced sensitivity to higher-order Zeeman shifts and atomic density shifts. This transition further supports degenerate two-photon Doppler-free spectroscopy due to an anomalously small $\approx\!87$ GHz detuning from the $6s6p\;^1P_1$ intermediate state. From our implementation of CI theory to the relatively low-lying $6s23s\;^1S_0$ Rydberg state, we predict state properties needed for estimating sensor performance including a relatively suppressed $C_6$ coefficient, a relatively long natural lifetime, and reduced second-order Zeeman sensitivity. We further extend the CI technique to the $6sns\;^1S_0$ states with $n=14$ to $n=26$ in the Appendix. We have also taken the first steps towards quantum sensing by performing degenerate two-photon Doppler-free spectroscopy of the $6s^2\;^1S_0\rightarrow6s23s\;^1S_0$ transition and demonstrated three distinct detection channels: trap-loss, lattice-trapped metastable state population, and fluorescence and have estimated the natural lifetime to be $1.8~\mu$s. Although demonstrated here with laser cooling, Doppler-free excitation with fluorescence detection is compatible with atomic vapor cells~\cite{lemke2019method, li_2026}. Similar to other multi-photon transitions, the $6s^2\;^1S_0 \rightarrow 6s23s\;^1S_0$ transition possesses a large probe ac Stark shift that if uncontrolled will limit the performance of a quantum sensor. However, multiple techniques can be applied to mitigate the probe Stark shifts~\cite{gerginov_2018, tofful_171yb_2024, sanner_2018}.

\begin{acknowledgments}
This work was supported by NIST, ONR, NSF QLCI Award OMA – 2016244. The authors thank Y. J. Wang and N. Schlossberger for their careful reading of the manuscript and critical feedback.
\end{acknowledgments}

\appendix

%


\clearpage 

\section{Configuration interaction calculations of the $6sns\;^1S_0$ states with $n=14$ to $n=26$}
\label{app: ci}

\label{sec: theory}
In the present work, we employ a configuration interaction (CI) method, treating Yb as a divalent system. The method, originally developed to evaluate properties of low-lying states of divalent systems~\cite{CheCheJoh93,SavJoh02}, is outlined in the following section. We regard extension to the Rydberg regime as a trial exercise, though with a well-motivated objective in mind. Namely, we aim to see if evaluated $C_6$ coefficients for the $6sns\,{^1S_0}$ states with $n\approx23$ exhibit suppression consistent with the extrapolated results of Ref.~\cite{VaiJonPot12}. Our approach is completely distinct from the single-active-electron model used in Ref.~\cite{VaiJonPot12}, providing an independent cross check against the extrapolated results. In addition to $C_6$ coefficients, we calculate other properties relevant for spectroscopy involving these Rydberg states, including lifetimes, polarizabilities, and BBR shifts. Available experimental data is limited to energies and lifetimes. Comparison with this data provides a general sense of the computational accuracy, though, due to a lack of experimental data for the other properties, we refrain from making quantitative assertions regarding the computational accuracy. We discuss, in different contexts throughout, implications of the weakly bound $4f$ electron shell in Yb.


Our computational method starts by generating single-electron orbitals as solutions to the relativistic Dirac equation. The potential is taken to be the Coulomb potential of the nucleus plus the Dirac-Hartree-Fock potential for the core electrons. We treat Yb as a divalent system, with the $4f$ electrons included in the core. The radial part of the orbitals is expanded in terms of $B$-splines~\cite{JohBluSap88,BacCorDec01}, using the dual-kinetic-balance prescription~\cite{ShaTupYer04,BelDer08}. The $B$-spline set has a finite radial extent, which is akin to confining the atom in a cavity. A sufficiently large cavity radius is required to faithfully represent states of the free atom. For instance, the $6sns\,{^1S_0}$ and $6snp\,{^{1,3}P_1}$ series each have a quantum defect of $\approx\!4$. Meanwhile, for a $29s$ or $29p$ hydrogenic orbit, the probability of finding the electron at radial distances $>\!2000a_B$ is $<\!10^{-4}$. This indicates that $2000a_B$ is an adequate cavity radius to represent the $6sns\,{^1S_0}$ and $6snp\,{^{1,3}P_1}$ states of Yb with $n$ up to $\approx\!33$, or even higher. For states higher in the spectrum, the presence of the cavity wall is felt, with the states of the cavity-confined atom no longer identifying with states of the free atom. Nevertheless, these states can be used in sum-over-states perturbation expressions, with the corresponding discrete set of states capable of capturing contributions from higher-lying states of the free atom, including continuum contributions.

There is a computational cost associated with larger cavity radii. Simply scaling the $B$-splines to a larger radius would imply a lower density of $B$-splines, compromising the representation of the radial wave functions. Thus, a larger cavity radius demands a larger number of $B$-splines. Also, the radial wave functions are ultimately projected onto a common radial grid, which is used for numerical integration prevalent in subsequent steps. A larger cavity radius demands more grid points and correspondingly more computational operations.

We reiterate that the potential used to generate the single-electron orbitals includes the Coulomb potential of the nucleus plus the Dirac-Hartree-Fock potential for the $N-2$ core electrons, where $N=70$ for Yb. With this potential, a valence electron ``sees'' a Coulomb potential with a charge of $+2e$, rather than $+e$, outside of the core region. As such, the generated valence orbitals are not expected to relate directly to Rydberg orbitals of the neutral atom. For instance, the $23s$ orbital obtained by solving the relativistic Dirac equation will be pulled much closer to the nucleus than the ``true'' $23s$ Rydberg orbital. However, the purpose of the single-electron orbitals is to construct a basis of many-body states for the subsequent steps, and this basis can be made sufficiently complete by including enough orbitals.

The many-body basis states are given in terms of Slater determinants. More specifically, they are linear combinations of Slater determinants satisfying a given symmetry $J^\pi$, where $\pi=\pm$ specifies parity and $J$ specifies total electronic angular momentum. We include configurations consisting of two valence electrons outside of the closed core. The many-body Hamiltonian includes all electron-nucleus and electron-electron Coulomb interactions. For a given symmetry, the many-body Hamiltonian is evaluated between the basis states, with the resulting matrix then being diagonalized. The eigenvalues and eigenstates are subsequently associated with energies and states of the (cavity-confined) atom. With these eigenstates, electric dipole or other matrix elements can be evaluated.

We note that the $4f$ electrons are relatively loosely bound. The Yb energy spectrum has several states with an unfilled $4f$ electron shell (referred to as $4f$-excited states throughout) residing below the first ionization threshold~\cite{NIST_ASD}. Such configurations are excluded from our basis, with the corresponding atomic states not being represented. Likewise, admixtures of these configurations are absent in our CI eigenstates. This is acceptable, provided the neglected configuration mixing is not too strong for the state of interest. For the $J^\pi=0^+$ energy spectrum, for instance, a $4f$-excited state resides between the $6s18s\,{^1S_0}$ and $6s19s\,{^1S_0}$ states, such that $4f$-excited admixtures within these two states may be appreciable. Meanwhile, the $6s23s\,{^1S_0}$ state is farther away in the spectrum, with the omitted $4f$-excited admixtures anticipated to be much smaller. Also, for properties involving a sum over states, neglect of configuration mixing in the space being summed over (e.g., $J^\pi=1^-$) can be acceptable, even if the mixing is relatively strong~\cite{DzuDer10}. While some effects attributed to core-excited configurations (e.g., self-energy) can be accounted for with many-body perturbation theory~\cite{SavJoh02}, we exclude these effects in this initial study. Such effects are known to be important for the low-lying atomic states, while here we are primarily interested in properties of the Rydberg states.

We test the stability of our results by performing calculations with two sets of computational parameters. Specifications for the two sets, designated $A$ and $B$, are provided in Table~\ref{Tab:bases}. In order to represent $6sns\,{^1S_0}$ states as well as states that couple to the $6sns\,{^1S_0}$ states through the electric dipole operator, the CI method described above is applied for both $J^\pi=0^+$ and $J^\pi=1^-$ symmetries.

\begin{table}[t]
\caption{Parameter sets for the CI method, labeled $A$ and $B$. Partial waves are understood to be relativistic partial waves (e.g., $p$ encapsulates both $p_{1/2}$ and $p_{3/2}$ partial waves). The ``orbitals'' list specifies the number of Dirac-Hartree-Fock orbitals used per partial wave in the construction of the many-body basis states. For instance, $125$ in the second entry of the list indicates that the lowest $125$ orbitals are used for both $p$ partial waves, inclusive of core orbitals.}
\label{Tab:bases}
\begin{ruledtabular}
\begin{tabular}{lcc}
							& $A$						& $B$					\\
\hline
\vspace{-3mm}\\
cavity radius ($a_B$)		& $2000$					& $2500$				\\
number of splines			& $400$						& $600$					\\
spline order				& $7$						& $7$					\\
partial waves				& $s,p,d,f,g,h$				& $s,p,d,f,g,h$			\\
number of grid points		& $2000$					& $3000$				\\
&	\multicolumn{2}{c}{$J^\pi=0^+$}												\\
number of basis states		& $26564$					& $29084$				\\
orbitals					& $120,120,40,$				& $160,120,40,$			\\
							& $40,40,40$				& $30,25,20$			\\
&	\multicolumn{2}{c}{$J^\pi=1^-$}												\\
number of basis states		& $29590$					& $29656$				\\
orbitals					& $90,90,30,$				& $70,125,30,$			\\
							& $30,30,30$				& $22,18,16$			
\end{tabular}
\end{ruledtabular}
\end{table}

\subsection{Energies}

Tables~\ref{Tab:Energies0e} and \ref{Tab:Energies1o} present energy spectra for the $J^\pi=0^+$ and $J^\pi=1^-$ symmetries, respectively, encompassing Rydberg states through $n=29$. Experimental and CI energies are given, including both CI parameter sets $A$ and $B$. For the experimental energies, we take the ground state to have zero energy, with other values taken from Refs.~\cite{NIST_ASD,LehZulMai18}. For the CI energies, we could likewise fix the ground state to have zero energy by adding the appropriate offset to the output CI eigenvalues. However, as we are more interested in Rydberg states in this work, we instead choose an offset that provides a more apparent comparison in this regime, with the same offset being applied for both symmetries. For concreteness, we fix the CI energy for the last state in Table~\ref{Tab:Energies0e}, the $6s29s\,{^1S_0}$ state, such that it coincides with the corresponding experimental value. The CI energies are simply listed in order, although we recall that the $4f$-excited states are not represented in our divalent CI approach. Thus, for appropriate comparison, we must skip over the $4f$-excited states of the experimental energy spectrum. We note that the NIST Atomic Structure Database~\cite{NIST_ASD} lists several $J^\pi=1^-$ states without an assigned configuration. We treat these as $4f$-excited states as well. By skipping the $4f$-excited states, we find respectable agreement between the CI and experimental energy spectra, especially in the Rydberg region of interest. Results are consistent for both parameter sets $A$ and $B$. Given the observed agreement with experiment, we can be confident in our assignment of the CI eigenstates. For instance, the twentieth CI eigenstate for the $J^\pi=0^+$ symmetry identifies with the $6s23s\,{^1S_0}$ state.

Tables~\ref{Tab:Energies0e} and \ref{Tab:Energies1o} demonstrate good agreement for energies of the Rydberg states with $n$ in the twenties, with absolute deviations between theory and experiment on the order of 0.1~cm$^{-1}$ across a range of states. Meanwhile, lower-lying states exhibit larger discrepancies. This is not surprising, with core-excited correlation effects not being accounted for in our approach, not even perturbatively, and with lower-lying states generally being more sensitive to effects of the core. Still, the maximum deviation of $\approx\!3700$~cm$^{-1}$ for the ground state represents just $\approx\!7\%$ of the ionization energy, which we deem satisfactory given our focus on the Rydberg states.

\begin{table*}[t]
\caption{First 28 states of Yb with $J^\pi=0^+$ symmetry. Table~\ref{Tab:Energies1o} displays the first 58 states of Yb with $J^\pi=1^-$ symmetry. In both cases, this encompasses Rydberg states associated with the first ionization threshold through $n=29$. The remainder of this caption serves both tables. $E_\mathrm{expt}$ and $E_\mathrm{CI}$ are experimental and CI energies, respectively (units of cm$^{-1}$, rounded to the nearest 0.1~cm$^{-1}$). Experimental energies are from Refs.~\cite{NIST_ASD,LehZulMai18,AymChaDel84} and have negligible uncertainty. These include $4f$-excited states, which are labeled here by electronic configuration only (when available). For the $J^\pi=1^-$ symmetry, the NIST Atomic Spectra Database lists several states without a configuration assignment, which we treat as $4f$-excited states as well. The CI results are given for both parameter sets $A$ and $B$, as specified in Table~\ref{Tab:bases}. In listing the CI energies, $4f$-excited states are skipped, as these states are not represented with our divalent CI method. For a more apparent comparison of the Rydberg energy spectra, the CI energy offset is chosen such that the CI and experimental energies for the $6s29s\,{^1S_0}$ state coincide. For the $J^\pi=1^-$ spectra, there are inconsistencies within the literature for the assignment of the spin quantum number for the $6snp\,{^{1,3}P_1}$ series~\cite{AymDebRob80,AymChaDel84,AliYasNad99,ZhaLar91,XinWanZha94,JiaLuoPen95}, while the NIST database uses $jj$-coupling instead of $LS$-coupling for $n\geq10$. Since such assignment is not critical for our purposes, we simply use $-$ and $+$ to denote the lower and upper state for each $n\geq10$. No experimental energy is available for the $6s9s\,{^1S_0}$ state.}
\label{Tab:Energies0e}
\begin{ruledtabular}
\begin{tabular}{cccccc}
&&\multicolumn{2}{c}{$A$}&\multicolumn{2}{c}{$B$}\\
state & $E_\mathrm{expt}$ 
& $E_\mathrm{CI}$ & $E_\mathrm{CI}-E_\mathrm{expt}$
& $E_\mathrm{CI}$ & $E_\mathrm{CI}-E_\mathrm{expt}$
\\
\hline
\vspace{-3mm}\\
$6s^2~{^1S_0}$	& $0$      	& $3709.3$	& $3709.3$	& $3774.4$	& $3774.4$	\\
$6s7s~{^1S_0}$	& $34350.6$	& $34980.1$	& $629.6$	& $35015.7$	& $665.1$	\\
$6s8s~{^1S_0}$	& $41939.8$	& $42290.0$	& $350.2$	& $42312.9$	& $373.1$	\\
$6p^2~{^3P_0}$	& $42436.8$	& $42400.6$	& $-36.2$	& $42423.5$	& $-13.4$	\\
$6s9s~{^1S_0}$	&          	& $45398.4$	&			& $45409.6$	&			\\
$4f^{13}5d6s6p$	& $46081.4$	&          	&			&			&			\\
$6s10s~{^1S_0}$	& $46893.1$	& $46980.3$	& $87.1$	& $46989.1$	& $95.9$	\\
$6s11s~{^1S_0}$	& $47808.4$	& $47859.6$	& $51.2$	& $47869.9$	& $61.6$	\\
$6p^2~{^1S_0}$	& $48344.3$	& $48325.4$	& $-18.9$	& $48335.7$	& $-8.6$	\\
$6s12s~{^1S_0}$	& $48723.4$	& $48693.3$	& $-30.1$	& $48696.5$	& $-26.9$	\\
$6s13s~{^1S_0}$	& $49046.3$	& $49035.5$	& $-10.8$	& $49036.7$	& $-9.6$	\\
$6s14s~{^1S_0}$	& $49302.4$	& $49299.9$	& $-2.5$	& $49300.7$	& $-1.8$	\\
$6s15s~{^1S_0}$	& $49499.1$	& $49499.1$	& $0.0$		& $49499.7$	& $0.5$		\\
$6s16s~{^1S_0}$	& $49650.0$	& $49651.3$	& $1.4$		& $49651.7$	& $1.8$		\\
$6s17s~{^1S_0}$	& $49767.7$	& $49769.8$	& $2.1$		& $49770.1$	& $2.4$		\\
$6s18s~{^1S_0}$	& $49859.4$	& $49863.7$	& $4.3$		& $49863.9$	& $4.5$		\\
$4f^{13}5d6s6p$	& $49897.2$	&          	&			&			&			\\
$6s19s~{^1S_0}$	& $49940.6$	& $49939.3$	& $-1.4$	& $49939.5$	& $-1.2$	\\
$6s20s~{^1S_0}$	& $50001.3$	& $50001.0$	& $-0.2$	& $50001.2$	& $-0.1$	\\
$6s21s~{^1S_0}$	& $50051.8$	& $50052.1$	& $0.3$		& $50052.2$	& $0.4$		\\
$6s22s~{^1S_0}$	& $50094.6$	& $50094.8$	& $0.2$		& $50094.9$	& $0.3$		\\
$6s23s~{^1S_0}$	& $50130.6$	& $50130.9$	& $0.2$		& $50130.9$	& $0.3$		\\
$6s24s~{^1S_0}$	& $50161.4$	& $50161.6$	& $0.2$		& $50161.7$	& $0.2$		\\
$6s25s~{^1S_0}$	& $50187.9$	& $50188.0$	& $0.2$		& $50188.1$	& $0.2$		\\
$6s26s~{^1S_0}$	& $50210.8$	& $50210.9$	& $0.1$		& $50210.9$	& $0.1$		\\
$6s27s~{^1S_0}$	& $50230.7$	& $50230.8$	& $0.1$		& $50230.8$	& $0.1$		\\
$6s28s~{^1S_0}$	& $50248.2$	& $50248.3$	& $0.0$		& $50248.3$	& $0.0$		\\
$6s29s~{^1S_0}$	& $50263.7$	& $50263.7$	& $0.0$		& $50263.7$	& $0.0$
\end{tabular}
\end{ruledtabular}
\end{table*}

\begin{table*}[t]
\caption{First 58 states of Yb with $J^\pi=1^-$ symmetry. Refer to the caption of Table~\ref{Tab:Energies0e} for details.}
\label{Tab:Energies1o}
\begin{ruledtabular}
\begin{tabular}{cccccc}
&&\multicolumn{2}{c}{$A$}&\multicolumn{2}{c}{$B$}\\
state & $E_\mathrm{expt}$ 
& $E_\mathrm{CI}$ & $E_\mathrm{CI}-E_\mathrm{expt}$
& $E_\mathrm{CI}$ & $E_\mathrm{CI}-E_\mathrm{expt}$
\\
\hline
\vspace{-3mm}\\
$6s6p~{^3P_1}$	& $17992.0$	& $18427.9$	& $435.9$	& $18434.4$	& $442.4$	\\
$6s6p~{^1P_1}$	& $25068.2$	& $27686.4$	& $2618.2$	& $27686.0$	& $2617.8$	\\
$4f^{13}5d6s^2$	& $28857.0$	&			& 			&			&			\\
				& $37414.6$	&			&			&			&			\\
$6s7p~{^3P_1}$	& $38174.2$	& $38386.0$	& $211.8$	& $38387.0$	& $212.8$	\\
				& $38422.4$	&			&			&			&			\\
$6s7p~{^1P_1}$	& $40564.0$	& $39373.4$	& $-1190.6$	& $39372.5$	& $-1191.5$	\\
				& $43532.8$	&			&			&			&			\\
$6s8p~{^3P_1}$	& $43659.4$	& $43743.1$	& $83.7$	& $43743.4$	& $84.0$	\\
$6s8p~{^1P_1}$	& $44017.6$	& $44009.2$	& $-8.4$	& $44008.7$	& $-8.9$	\\
				& $45181.7$	&			&			&			&			\\
$6s9p~{^3P_1}$	& $46078.9$	& $46146.6$	& $67.7$	& $46146.7$	& $67.7$	\\
				& $46174.2$	&			& 			&			&			\\ 
$6s9p~{^1P_1}$	& $46370.3$	& $46254.5$	& $-115.8$	& $46254.3$	& $-116.0$	\\
				& $47332.3$ &			&			&			&			\\
$6s10p~{-}$		& $47420.9$	& $47447.6$	& $26.7$	& $47447.6$	& $26.6$	\\
$6s10p~{+}$		& $47498.8$	& $47504.9$	& $6.0$		& $47504.8$	& $5.9$		\\
				& $47859.3$	& 			&			&			&			\\ 
$6s11p~{-}$		& $48212.1$	& $48234.1$	& $22.0$	& $48234.1$	& $22.0$	\\
$6s11p~{+}$		& $48258.5$	& $48270.2$	& $11.7$	& $48270.1$	& $11.6$	\\
$6s12p~{-}$		& $48719.2$	& $48746.7$	& $27.5$	& $48746.6$	& $27.4$	\\
$6s12p~{+}$		& $48761.8$	& $48771.7$	& $9.9$		& $48771.6$	& $9.8$		\\
$4f^{13}5d^26s$	& $49005.9$	&			&			&			&			\\
$6s13p~{-}$		& $49110.1$	& $49099.4$	& $-10.7$	& $49099.3$	& $-10.7$	\\
$6s13p~{+}$		& $49127.4$	& $49117.9$	& $-9.5$	& $49117.8$	& $-9.6$	\\
$6s14p~{-}$		& $49352.5$	& $49352.5$	& $0.0$		& $49352.5$	& $0.0$		\\
$6s14p~{+}$		& $49360.6$	& $49366.7$	& $6.1$		& $49366.7$	& $6.1$		\\
$6s15p~{-}$		& $49536.4$	& $49540.3$	& $3.8$		& $49540.2$	& $3.8$		\\
$6s15p~{+}$		& $49546.6$	& $49551.6$	& $5.1$		& $49551.6$	& $5.0$		\\
$6s16p~{-}$		& $49677.6$	& $49683.4$	& $5.8$		& $49683.4$	& $5.8$		\\
$6s16p~{+}$		& $49688.6$	& $49692.6$	& $4.1$		& $49692.6$	& $4.1$		\\
$6s17p~{-}$		& $49786.4$	& $49795.0$	& $8.6$		& $49795.0$	& $8.6$		\\
$6s17p~{+}$		& $49799.2$	& $49802.7$	& $3.5$		& $49802.6$	& $3.4$		\\
$6s18p~{-}$		& $49867.1$	& $49883.7$	& $16.6$	& $49883.7$	& $16.5$	\\
$6s18p~{+}$		& $49886.9$	& $49890.2$	& $3.3$		& $49890.2$	& $3.2$		\\
$4f^{13}5d^26s$	& $49920.0$	&			&			&			&			\\
$6s19p~{-}$		& $49957.9$	& $49955.4$	& $-2.5$	& $49955.3$	& $-2.5$	\\
$6s19p~{+}$		& $49969.2$	& $49960.9$	& $-8.2$	& $49960.9$	& $-8.3$	\\
$6s20p~{-}$		& $50016.0$	& $50014.1$	& $-1.9$	& $50014.0$	& $-1.9$	\\
$6s20p~{+}$		& $50021.3$	& $50018.9$	& $-2.4$	& $50018.9$	& $-2.4$	\\
$6s21p~{-}$		& $50064.1$	& $50062.8$	& $-1.4$	& $50062.7$	& $-1.4$	\\
$6s21p~{+}$		& $50067.8$	& $50067.0$	& $-0.7$	& $50067.0$	& $-0.8$	\\
$6s22p~{-}$		& $50104.6$	& $50103.6$	& $-1.0$	& $50103.6$	& $-1.1$	\\
$6s22p~{+}$		& $50107.6$	& $50107.4$	& $-0.3$	& $50107.3$	& $-0.3$	\\
$6s23p~{-}$		& $50139.0$	& $50138.2$	& $-0.8$	& $50138.2$	& $-0.8$	\\
$6s23p~{+}$		& $50141.6$	& $50141.6$	& $0.0$		& $50141.5$	& $-0.1$	\\
$6s24p~{-}$		& $50168.4$	& $50167.8$	& $-0.6$	& $50167.8$	& $-0.6$	\\
$6s24p~{+}$		& $50170.7$	& $50170.8$	& $0.1$		& $50170.7$	& $0.0$		\\
$6s25p~{-}$		& $50193.7$	& $50193.2$	& $-0.4$	& $50193.2$	& $-0.5$	\\
$6s25p~{+}$		& $50195.9$	& $50196.0$	& $0.0$		& $50195.9$	& $0.0$		\\
$6s26p~{-}$		& $50215.6$	& $50215.3$	& $-0.3$	& $50215.3$	& $-0.4$	\\
$6s26p~{+}$		& $50217.7$	& $50217.8$	& $0.1$		& $50217.7$	& $0.0$		\\
$6s27p~{-}$		& $50234.9$	& $50234.6$	& $-0.3$	& $50234.6$	& $-0.3$	\\
$6s27p~{+}$		& $50236.9$	& $50236.8$	& $0.0$		& $50236.8$	& $-0.1$	\\
$6s28p~{-}$		& $50251.7$	& $50251.5$	& $-0.2$	& $50251.5$	& $-0.3$	\\
$6s28p~{+}$		& $50253.7$	& $50253.6$	& $-0.1$	& $50253.5$ & $-0.1$	\\
$6s29p~{-}$		& $50266.7$	& $50266.5$	& $-0.2$	& $50266.4$	& $-0.2$	\\
$6s29p~{+}$		& $50268.5$	& $50268.4$	& $-0.1$	& $50268.3$	& $-0.2$
\end{tabular}
\end{ruledtabular}
\end{table*}
\subsection{Lifetimes}

The radiative decay rate for a state with $J^\pi=0^+$ symmetry is given by
\begin{gather}
\gamma_\mathrm{nat}=\frac{1}{3\pi\epsilon_0\hbar c^3}\sum_k\omega_{rk}^3\left|\langle r||\mathbf{D}||k\rangle\right|^2,
\label{Eq:gammarad}
\end{gather}
where $r$ labels the $J^\pi=0^+$ state of interest and $k$ runs over all $J^\pi=1^-$ states that are energetically lower. Here $\omega_{rk}=\left(E_r-E_k\right)/\hbar$, with $E_i$ the energy of state $i$, and $\langle r||\mathbf{D}||k\rangle$ is the reduced matrix element of the electric dipole operator. The summation over $k$ is understood to exclude summation over magnetic quantum numbers. The natural lifetime of the state is $1/\gamma_\mathrm{nat}$.

We obtain {\it ab initio} results for the natural lifetimes by combining CI matrix elements with CI energies. One might expect improved results by replacing the CI energies with experimental energies, as the experimental energies have negligible uncertainty. However, there may be more favorable combinations of theoretical and experimental data. Consider oscillator strengths, proportional to the product $\omega_{rk}\left|\langle r||\mathbf{D}||k\rangle\right|^2$. In the nonrelativistic limit, the Thomas-Reiche-Kuhn (TRK) sum rule insists that oscillator strengths sum to a constant value. This motivates us to combine CI results for the product $\omega_{rk}\left|\langle r||\mathbf{D}||k\rangle\right|^2$ together with experimental energies (for remaining $\omega_{rk}$) in favor of CI matrix elements with experimental energies. Results obtained using this prescription are referred to as ``modified CI'' results to distinguish them from {\it ab initio} CI results.

We recall that the $4f$-excited states are not represented in our divalent CI approach, such that decay to these states is not formally taken into account. These decay channels are nominally forbidden from configuration considerations, but can be opened up through configuration mixing. This is an example of where neglect of configuration mixing, even relatively strong configuration mixing, can be acceptable. In the $J^\pi=1^-$ spectrum, for instance, there is known to be appreciable mixing between the $6s6p\,{^1P_1}$ state and the $4f$-excited state lying immediately above it (with respective energies of 25068.2~cm$^{-1}$ and 28857.0~cm$^{-1}$; see Table~\ref{Tab:Energies1o}). This mixing opens up decay to the latter state, while reducing decay to the former state. The overall decay rate is only expected to be modestly affected by the mixing between these states. This is analogous to calculations of the polarizabilities and $C_6$ coefficients for the clock states of Yb performed in Ref.~\cite{DzuDer10}, where neglect of configuration mixing in the intermediate-state space was deemed acceptable.

\begin{figure}[t]
\includegraphics[width=246pt]{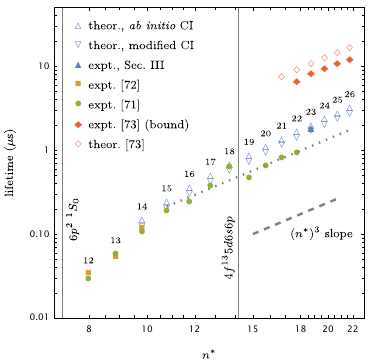}
\caption{Theoretical and experimental natural lifetimes for $6sns\,{^1S_0}$ states of Yb. The first three data sets listed in the legend are from the present work, while the others are from the literature. See Table~\ref{Tab:lifetimes} for the complete data. The ``bound'' designation in the legend indicates that the results are interpreted as a lower bound on the lifetime. Numbers identify the principal quantum number $n$. The lifetimes are plotted versus effective principal quantum number $n^*$, as determined from experimental energies~\cite{LehZulMai18}. Note that $n^*$ is essentially a nonlinear energy scale; vertical lines indicate the location of other $J^\pi=0^+$ states on this energy scale. This includes the $4f$-excited state residing between the $6s18s\,{^1S_0}$ and $6s19s\,{^1S_0}$ states that perturbs the lifetimes in this vicinity. The dashed line depicts the slope, on the log-log plot, associated with the nominal $\left(n^*\right)^3$ scaling for lifetimes of a given Rydberg series. The dotted line has this slope, but is taken to pass through the last ($n=22$) data point from Ref.~\cite{DadWanJia87}.}
\label{Fig:lifetime}
\end{figure}

Literature values are available for the lifetimes of $6sns\,{^1S_0}$ states, to which our CI results can be compared. Table~\ref{Tab:lifetimes} presents theoretical and experimental results for these lifetimes. We include both {\it ab initio} CI and modified CI results. In each case, the set-$A$ and set-$B$ results are consistent to better than one percent, with only the set-$A$ results being given in Table~\ref{Tab:lifetimes}. Experimental values are taken from three literature sources, Refs.~\cite{BauBraMai87,DadWanJia87,FanXieZha01}, which cover different ranges of $n$ with some overlap among them. For Refs.~\cite{BauBraMai87,DadWanJia87}, the lifetimes are seen to agree for the three states in common ($n=12,13,14$). Reference~\cite{FanXieZha01} provides both experimental and theoretical results. The authors note that BBR-induced downward transitions could lead to observed lifetimes smaller than the natural lifetimes, while estimating this effect to be at the $\sim\!10\%$ level. We thus take the reported lifetimes from Ref.~\cite{FanXieZha01} as lower bounds on the natural lifetimes. In any case, there is significant tension between Refs.~\cite{DadWanJia87} and \cite{FanXieZha01} for the single state in common ($n=22$).
We also include the lifetime for the $6s23s\,{^1S_0}$ state inferred from spectroscopic linewidth measurements described in Sec.~\ref{sec: experiment} of this work.

\newcommand{\s}[1]{\ensuremath{6s{#1}s\,{^1S_0}}}

\begin{table*}[t]
\caption{Natural lifetimes for the $6sns\,{^1S_0}$ states of Yb, for $12\leq n\leq26$. The first three data columns give results from the present work. The experimental result is inferred from the measured linewidth described in Sec.~\ref{sec: experiment}. The remaining data columns give results from the literature. For the experimental results of Ref.~\cite{FanXieZha01}, the uncertainty is estimated to be up to $\sim\!8\%$, though discrepancy with Ref.~\cite{DadWanJia87} for the $6s22s~{^1S_0}$ state far exceeds that estimate. The units are microseconds.}

\label{Tab:lifetimes}
\begin{ruledtabular}
\begin{tabular}{lcccccccc}
& \multicolumn{3}{c}{present}	& \multicolumn{4}{c}{literature}	\\
\cline{2-4}\cline{5-8}

\vspace{-3mm}\\
state
& theor.\footnotemark[1]	& theor.\footnotemark[2] & expt.\footnotemark[3]
& expt.~\cite{BauBraMai87}	& expt.~\cite{DadWanJia87} 
& expt.~\cite{FanXieZha01}	& theor.~\cite{FanXieZha01}
\\
\hline
\vspace{-3mm}\\

\s{12}	&			& 			&	           & $0.0336(50)$	& $0.0292(15)$	&				&			\\
\s{13}	&			&			&	           & $0.0531(79)$	& $0.0583(29)$	&				&			\\
\s{14}	& $0.147$	& $0.123$	&	           & $0.115(17)$	& $0.1064(32)$	&				&			\\
\s{15}	& $0.242$	& $0.203$	&	           &				& $0.1909(75)$	&				&			\\
\s{16}	& $0.360$	& $0.304$	&	           &				& $0.2438(90)$	&				&			\\
\s{17}	& $0.502$	& $0.426$	&	           &				& $0.377(20)$	&				&			\\
\s{18}	& $0.671$	& $0.571$	&	           &				& $0.632(32)$	&				&			\\
\s{19}	& $0.867$	& $0.737$	&	           &				& $0.466(18)$	&				&			\\
\s{20}	& $1.09$	& $0.935$	&	           &				& $0.648(21)$	&				&			\\
\s{21}	& $1.35$	& $1.16$	&	           &				& $0.815(32)$	&			    &$7.6$		\\
\s{22}	& $1.65$	& $1.41$	&	           &				& $0.946(38)$	& $\geq6.5$	    &$9.1$	    \\
\s{23}	& $1.98$	& $1.70$	& $1.8$        &                &  				& $\geq8.1$		& $10.7$	\\
\s{24}	& $2.35$	& $2.02$	&	           &				& 				& $\geq9.4$		& $12.5$	\\
\s{25}	& $2.76$	& $2.37$	&	           &				&				& $\geq10.6$	& $14.5$	\\
\s{26}	& $3.21$	& $2.76$	&	           &				& 				& $\geq12.1$	& $16.7$    \\
\end{tabular}
\end{ruledtabular}
\footnotetext[1]{{\it ab initio} CI}
\footnotetext[2]{modified CI}
\footnotetext[3]{Sec.~\ref{sec: experiment}}
\end{table*}

In Figure~\ref{Fig:lifetime}, we plot the lifetime data from Table~\ref{Tab:lifetimes} versus effective principal quantum number $n^*$, displayed on a log-log plot. Lifetimes for a given Rydberg series nominally scale as $\left(n^*\right)^3$. Focusing initially on the experimental results from Ref.~\cite{DadWanJia87}, we see that the $\left(n^*\right)^3$ scaling generally holds for $n\gtrsim 15$ (refer to the dotted line in Fig.~\ref{Fig:lifetime}), except for notable deviations in the vicinity of $n=18,19$. Such deviations are not surprising, given the presence of the $4f$-excited ``perturber'' state in this region of the $J^\pi=0^+$ energy spectrum (refer to the vertical gray line in Fig.~\ref{Fig:lifetime}). As there are no known perturber states higher in the Rydberg spectrum, it is reasonable to expect that lifetimes for states with $n>22$ may be approximated through extrapolation of this experimental data (e.g., following the dotted line in Fig.~\ref{Fig:lifetime}). Next, we focus on the CI results, simultaneously considering the {\it ab initio} CI and modified CI results. For the last several states, the CI results exhibit an approximate scaling of $\left(n^*\right)^{3.4}$. Notably, no deviations occur in the vicinity of the perturber state. This is not surprising either, given that the $4f$-excited states are excluded in the divalent CI approach. Ignorant of any other data, comparison of the CI results with the experimental results of Ref.~\cite{DadWanJia87} would lead us to conclude that the CI results overestimate the lifetimes in the region of interest (i.e., $n\approx23$). More specifically, the {\it ab initio} CI and modified CI approaches appear to overestimate the lifetimes by respective amounts of $\approx70\%$ and $\approx50\%$. However, the \s{23} lifetime inferred from the measured linewidth in this work (see Sec.~\ref{sec: experiment}) is compatible with the CI results but is incompatible with extrapolation of the Ref.~\cite{DadWanJia87} data. Experimental and theoretical results from Ref.~\cite{FanXieZha01} further cloud the picture, suggesting substantially larger lifetimes. Overall, the CI results demonstrate reasonable agreement with the available lifetime data, to the extent that the available data allows us to say as much.

\subsection{$C_6$ coefficients, dc polarizabilities, BBR shifts, and BBR depopulation rates}

The $C_6$ coefficient, the dc polarizability $\alpha_\mathrm{dc}$, the BBR shift $\delta E_\mathrm{BBR}$, and the BBR depopulation rate $\gamma_\mathrm{BBR}$ are given by~\cite{PorSafDer14,FarWin81}
\begin{gather*}
C_6=-\frac{1}{24\pi^2\epsilon_0^2\hbar}\sum_{kl}
\frac{\left|\langle r||\mathbf{D}||k\rangle\right|^2\left|\langle r||\mathbf{D}||l\rangle\right|^2}{\omega_{rk}+\omega_{rl}},
\\
\alpha_\mathrm{dc}=-\frac{2}{3\hbar}\sum_{k}
\frac{\left|\langle r||\mathbf{D}||k\rangle\right|^2}{\omega_{rk}},
\\
\delta E_\mathrm{BBR}=\frac{\left(k_BT\right)^3}{6\pi^2\epsilon_0\hbar^3c^3}
\sum_k\left|\langle r||\mathbf{D}||k\rangle\right|^2F\left(\frac{\hbar\omega_{rk}}{k_BT}\right),
\\
\gamma_\mathrm{BBR}=\frac{1}{3\pi\epsilon_0\hbar c^3}
\sum_k\left|\langle r||\mathbf{D}||k\rangle\right|^2
\frac{\left|\omega_{rk}\right|^3}{\exp\left(\displaystyle\frac{\hbar\left|\omega_{rk}\right|}{k_BT}\right)-1}.
\end{gather*}
Here $F(y)$ is the Farley-Wing function,
\begin{gather*}
F(y)=\int_0^\infty\left(\frac{1}{y+x}+\frac{1}{y-x}\right)\frac{x^3}{e^x-1}dx,
\end{gather*}
where the Cauchy principal value is taken for the integral. In contrast to Eq.~(\ref{Eq:gammarad}) above, the sum over states in these expressions runs over all $J^\pi=1^-$ states, not just those energetically below the $J^\pi=0^+$ state under consideration. Again, the summations exclude magnetic quantum numbers. The rate $\gamma_\mathrm{BBR}$ is associated with transitions out of the state driven by the BBR field, reducing the effective lifetime of the state.

\begin{table*}[t]
\caption{Properties of the $6sns\,{^1S_0}$ states of Yb calculated with the CI method, for $14\leq n\leq26$. The BBR shift $\delta E_\mathrm{BBR}$, the slope of the BBR shift with respect to temperature $S_\mathrm{BBR}\equiv\partial\left(\delta E_\mathrm{BBR}\right)/\partial T$, and the BBR depopulation rate $\gamma_\mathrm{BBR}$ are evaluated for a BBR temperature of 300~K.}
\label{Tab:otherprops}
\begin{ruledtabular}
\begin{tabular}{lcccccccccc}
state
& \multicolumn{2}{c}{$C_6$~~($\times10^{12}$~a.u.)}
& \multicolumn{2}{c}{$\alpha_\mathrm{dc}$~~($\times10^{8}$~a.u.)}
& \multicolumn{2}{c}{$\delta E_\mathrm{BBR}/h$~~(kHz)}
& \multicolumn{2}{c}{$S_\mathrm{BBR}/h$~~(Hz/K)}
& \multicolumn{2}{c}{$\gamma_\mathrm{BBR}$~~(ms$^{-1}$)}	\\
\cline{2-3}\cline{4-5}\cline{6-7}\cline{8-9}\cline{10-11}
\vspace{-3mm}\\
& CI\footnotemark[1] & CI\footnotemark[2]
& CI\footnotemark[1] & CI\footnotemark[2]
& CI\footnotemark[1] & CI\footnotemark[2]
& CI\footnotemark[1] & CI\footnotemark[2]
& CI\footnotemark[1] & CI\footnotemark[2]
\\
\hline
\vspace{-3mm}\\
\s{14}	& $0.273$	& $0.384$	& $0.439$	& $0.527$	& $4.32$	& $4.31$
		& $22.9$	& $22.9$	& $64.3$	& $61.5$	\\
\s{15}	& $0.640$	& $1.06$	& $0.809$	& $1.00$	& $4.13$	& $4.23$
		& $21.8$	& $22.0$	& $63.9$	& $62.5$	\\
\s{16}	& $1.47$	& $2.73$	& $1.45$	& $1.88$	& $3.91$	& $4.01$
		& $20.9$	& $21.1$	& $60.8$	& $59.8$	\\
\s{17}	& $3.25$	& $9.18$	& $2.51$	& $4.03$	& $3.72$	& $3.82$
		& $20.2$	& $20.5$	& $56.9$	& $55.5$	\\
\s{18}	& $6.95$	& $112$		& $4.22$	& $20.4$	& $3.56$	& $3.66$
		& $19.8$	& $20.1$	& $52.8$	& $49.9$	\\
\s{19}	& $14.4$	& $9.07$	& $6.86$	& $5.23$	& $3.42$	& $3.54$
		& $19.4$	& $19.6$	& $48.8$	& $55.8$	\\
\s{20}	& $28.8$	& $-17.1$	& $10.9$	& $7.98$	& $3.31$	& $3.24$
		& $19.0$	& $18.9$	& $45.1$	& $44.4$	\\
\s{21}	& $56.1$	& $2.22$	& $16.8$	& $12.9$	& $3.22$	& $3.16$
		& $18.7$	& $18.6$	& $41.6$	& $41.3$	\\
\s{22}	& $106$		& $34.1$	& $25.4$	& $20.6$	& $3.13$	& $3.10$
		& $18.5$	& $18.4$	& $38.5$	& $38.3$	\\
\s{23}	& $197$		& $76.4$	& $37.6$	& $30.8$	& $3.06$	& $3.04$
		& $18.3$	& $18.2$	& $35.6$	& $35.5$	\\
\s{24}	& $355$		& $168$		& $54.7$	& $45.6$	& $3.00$	& $2.98$
		& $18.1$	& $18.1$	& $33.0$	& $33.0$	\\
\s{25}	& $629$		& $345$		& $78.3$	& $66.8$	& $2.95$	& $2.94$
		& $17.9$	& $17.9$	& $30.7$	& $30.7$	\\
\s{26}	& $1090$	& $681$		& $111$		& $96.5$	& $2.91$	& $2.90$
		& $17.8$	& $17.8$	& $28.6$	& $28.6$	
\end{tabular}
\end{ruledtabular}
\footnotetext[1]{{\it ab initio} CI}
\footnotetext[2]{modified CI}
\end{table*}
We calculate the properties $C_6$, $\alpha_\mathrm{dc}$, $\delta E_\mathrm{BBR}$, and $\gamma_\mathrm{BBR}$ for the $6sns\,{^1S_0}$ states of Yb with $14\leq n\leq26$. We also calculate $S_\mathrm{BBR}\equiv\partial\left(\delta E_\mathrm{BBR}\right)/\partial T$ by taking a numerical derivative of $\delta E_\mathrm{BBR}$ with respect to temperature. The BBR-related properties $\delta E_\mathrm{BBR}$, $S_\mathrm{BBR}$, and $\gamma_\mathrm{BBR}$ are calculated assuming a BBR temperature of 300~K. For the modified CI calculations, experimental energies are used for the $6sns\,{^1S_0}$ state under consideration, as well as $J^\pi=1^-$ intermediate states through Rydberg states with $n=29$ (the last line entry in Table~\ref{Tab:Energies1o}). For intermediate states above this, where the cavity wall begins to influence the spectrum, we use the CI energies. The results are compiled in Table~\ref{Tab:otherprops}. Both {\it ab initio} CI and modified CI results are presented. Only the set-$A$ results are tabulated. The set-$A$ and set-$B$ results are consistent to within a few percent, with the following exceptions for the $C_6$ coefficients. For the {\it ab initio} CI calculations, the set-$B$ results are 5\% to 9\% higher than the set-$A$ results. For the modified CI calculations, the $C_6$ coefficient for the \s{21} state is accidentally small, with the set-$B$ result being $14\%$ larger than the set-$A$ result.

For the $C_6$ coefficients, there is significant cancellation between positive and negative contributions. Using the modified CI set-$A$ result for the $6s23s\,{^1S_0}$ state as an example, we find cumulative positive and negative contributions of $3.88\times10^{14}$ a.u.\ and $-3.12\times10^{14}$ a.u., respectively, yielding a net result of $7.6\times10^{13}$ a.u. (a.u.\ stands for atomic units). The modified CI result for the $6s21s\,{^1S_0}$ state is accidentally small due to an even higher degree of cancellation.

\begin{figure}[H]
\includegraphics[width=246pt]{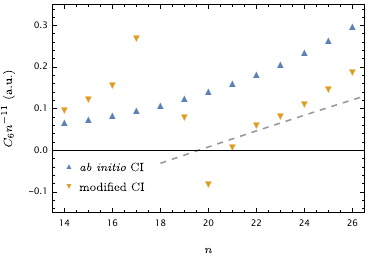}
\caption{$C_6$ coefficients for the $6sns\,{^1S_0}$ states of Yb calculated with the divalent CI method, for $14\leq n\leq26$ (data points). The $6s18s\,{^1S_0}$ modified CI result is out of the plot range. See Table \ref{Tab:otherprops} for the complete data. The dashed curve is the polynomial from Vaillant~{\it et al.}~\cite{VaiJonPot12}, extrapolated to lower values of $n$.}
\label{Fig:c6dataplot}
\end{figure}

Figure~\ref{Fig:c6dataplot} plots the $C_6$ coefficients from Table~\ref{Tab:otherprops} versus principal quantum number $n$. Generally speaking, we see order-of-magnitude agreement between the {\it ab initio} CI and modified CI results. The {\it ab initio} CI results exhibit smooth functional behavior with respect to $n$, while this is not the case for the modified CI results in the vicinity of $n=18,19$ (note that the $n=18$ result is out of the plot range). Evidently, the modified CI results ``feel'' the influence of $4f$-excited perturber state, despite the absence of such states from the divalent CI computational approach. This is in contrast to the lifetimes, which can be understood as follows. For radiative decay, the largest branching factor is to the $6s6p\,{^1P_1}$ state. Perturbations to the $6sns\,{^1S_0}$ energies caused by the perturber state amount to a relatively small effect on the $6sns\,{^1S_0}\rightarrow6s6p\,{^1P_1}$ transition frequency. Thus, replacing CI energies with experimental energies leads to a relatively small effect, primarily attributed to correction of the $6s6p\,{^1P_1}$ energy. For the $C_6$ coefficients, on the other hand, the dominant intermediate-state contributions are from neighboring Rydberg states with small transition frequencies. In this case, small changes in the Rydberg-state energies can have a relatively large effect on the transition frequencies for the dominant contributions. We note that, in addition to the $4f$-excited perturber state in the $J^\pi=0^+$ spectrum (at 49897.2~cm$^{-1}$), there is an energetically similar $4f$-excited perturber state in the $J^\pi=1^-$ spectrum (at 49920.0~cm$^{-1}$). By perturbing their respective Rydberg energy spectra, both perturber states influence the relevant transition frequencies. Figure~\ref{Fig:c6dataplot} also plots the polynomial fit to semi-empirical data from Vaillant~{\it et~al.}~\cite{VaiJonPot12}, extrapolated to lower values of $n$. Generally speaking, we see order-of-magnitude agreement between the CI results and the extrapolated polynomial of Ref.~\cite{VaiJonPot12}. Thus, the CI results provide additional evidence that the $C_6$ coefficients are highly suppressed for the $6sns\,{^1S_0}$ states of Yb with $n\approx23$. Table ~\ref{Tab: rydberg_props} presents atomic properties relevant to ac electrometry, including matrix elements and rf detection frequencies.

\begin{table*}[h]
\caption{Atomic properties relevant to ac electrometry. Reduced dipole matrix elements squared ($\left|\langle r||\mathbf{D}||k\rangle\right|^2$) between the $6s23s\;^1S_0$ state and the $6snp~J=1$ Rydberg states for $21\leq n\leq25$. The matrix elements are calculated with the CI theory. The rf detection frequencies ($E/h$) are calculated using the state energies given in Ref.~\cite{NIST_ASD}. The $6s23s\;^1S_0$ state energy is taken from Ref.~\cite{LehZulMai18}.}
\label{Tab: rydberg_props}
\begin{ruledtabular}
\begin{tabular}{lccc}
Rydberg State 
& $\left|\langle r||\mathbf{D}||k\rangle\right|^2$ (a.u.) \footnotemark[1]
& $\left|\langle r||\mathbf{D}||k\rangle\right|^2$ (a.u.) \footnotemark[2]
& $|E/h|$ (THz)
\\
\hline
\vspace{-3mm}\\
$6s_{1/2}21p_{1/2}$ & 37.882   & 37.860  & 1.996\\
$6s_{1/2}21p_{3/2}$ & 38.275  & 38.208   & 1.888 \\
$6s_{1/2}22p_{1/2}$ & 158.347  & 157.516    & 0.784 \\
$6s_{1/2}22p_{3/2}$ & 192.842  & 191.211    & 0.693 \\
$6s_{1/2}23p_{1/2}$ & 362.649  & 363.886   & 0.246 \\
$6s_{1/2}23p_{3/2}$ & 333.444   & 333.756   & 0.327 \\
$6s_{1/2}24p_{1/2}$ & 17.633   & 18.450   & 1.128\\
$6s_{1/2}24p_{3/2}$ & 10.002  & 9.246    & 1.198 \\
$6s_{1/2}25p_{1/2}$ & 10.288  & 10.596    & 1.885 \\
$6s_{1/2}25p_{3/2}$ & 1.810   & 1.474   & 1.953\\
\end{tabular}
\end{ruledtabular}
\footnotetext[1]{{\it ab initio} CI}
\footnotetext[2]{modified CI}
\end{table*}

\subsection{Additional considerations for BBR thermometry using Rydberg level shifts}
To extract an accurate temperature measurement using the measured transition frequency, knowledge of the transition frequency versus temperature is required, enabling conversion from one to the other. Note that even if theory could provide a sufficiently reliable result for the BBR frequency shift versus temperature, the unperturbed transition frequency would still be required, which theory is hardly capable of providing at a sufficient level. Inevitably, this implies that an atomic ``calibration'' measurement would be necessary, with the transition frequency being measured in a well-characterized BBR environment. The resulting calibration curve could then be employed to evaluate the BBR temperature in systems with more relaxed environmental control, with the calibration uncertainty ultimately setting a limit on the extracted temperature uncertainty. Our Yb OLCs use thermal shields to furnish the atoms with a well-characterized BBR environment~\cite{BelHinPhi14,hassan_cryo_2025}. Our latest version is capable of operating over a wide range of temperatures while maintaining excellent environmental control; this shield could potentially be used to obtain the requisite atomic calibration curve.

Reference~\cite{OvsDerGib11} proposed using Rydberg spectroscopy to measure the thermal radiation environment in an optical lattice clock, allowing for in-situ characterization of the BBR Stark shift to the clock transition. Whether that is done with Rydberg atoms trapped in the optical lattice, as originally proposed in Ref.~\cite{OvsDerGib11}, or with Doppler-free spectroscopy in free-space, as explored here, the spectral dependence of the underlying ac Stark shift must be considered. For the clock states, the spectral dependence manifests as an approximate $T^4$ scaling of the BBR shift~\cite{ItaLewWin82,PorDer06}. For the Rydberg states, the spectral dependence manifests as an approximate $T^2$ scaling of the BBR shift, as given by the universal expression in Sec.~\ref{Sec:applications}. That is, the clock frequency scales predominantly as $T^4$, while the ``sense'' frequency (i.e., the transition frequency between a low-lying state and a Rydberg state used for BBR sensing) scales predominantly as $T^2$. This difference could have practical implications if deviations from a Planckian spectral distribution are present. To illustrate this, we consider a hypothetical clock system in which the atoms are enclosed by two blackbody surfaces. We take the two surfaces to have temperatures $T_1$ and $T_2$, while subtending respective fractional solid angles $\rho_1$ and $\rho_2$ from the perspective of the atoms ($\rho_1+\rho_2=1$). In this case, the $T^4$ dependence for the clock frequency is simply replaced with $\left(\rho_1T_1^4+\rho_2T_2^4\right)$, while the $T^2$ dependence for the sense frequency is simply replaced with $\left(\rho_1T_1^2+\rho_2T_2^2\right)$. It follows that the BBR clock shift, as inferred from the sense frequency, incurs a fractional error of 
\begin{equation}
\frac{\left(\rho_1T_1^2+\rho_2T_2^2\right)^2}{\rho_1T_1^4+\rho_2T_2^4}-1.
\end{equation}
To make this example quantitative, we suppose $\rho_1=\rho_2=0.5$, $T_1=300$~K, and $T_2=305$~K, with the expression above evaluating to $-3\times10^{-4}$. For Yb, this translates to a fractional clock error of $7\times10^{-19}$. This example suggests that there is a fair amount of leeway in using the sense frequency to infer the BBR clock shift, despite the underlying differences in spectral response. Nevertheless, error stemming from a non-Planckian spectral distribution would need to be considered on a case-by-case basis. Note that here we have only considered the dominant scaling for the respective shifts, though smaller corrections (e.g., the ``dynamic'' correction to the BBR clock shift~\cite{ItaLewWin82,PorDer06}) would also need to be considered in a careful analysis.

\end{document}